\documentclass[twocolumn,showpacs,preprintnumbers,amsmath,amssymb]{revtex4}

\usepackage{graphicx}
\usepackage{dcolumn}
\usepackage{bm}

\begin{document}

\title{Demonstration of three-photon de Broglie wavelength
by projection measurement}

\author{B. H. Liu$^1$, F. W. Sun$^{1}$, Y. X. Gong$^{1}$, Y. F. Huang$^{1}$, Z. Y. Ou$^{1,2}\footnote{E-mail:
zou@iupui.edu}$, and G. C. Guo$^1$}
 \affiliation{$^1$Key Laboratory of Quantum Information,
 University of Science and Technology of China, \\CAS, Hefei, 230026, the People's Republic of China
 \\$^2$Department of Physics, Indiana
University-Purdue University Indianapolis, 402 N. Blackford
Street, Indianapolis, IN 46202}

\begin{abstract}
Two schemes of projection measurement are realized experimentally
to demonstrate the de Broglie wavelength of three photons without
the need for a maximally entangled three-photon state (the NOON
state). The first scheme is based on the proposal by Wang and
Kobayashi (Phys. Rev. A {\bf 71}, 021802) that utilizes a couple
of asymmetric beam splitters while the second one applies the
general method of NOON state projection measurement to
three-photon case. Quantum interference of three photons is
responsible for projecting out the unwanted states, leaving only
the NOON state contribution in these schemes of projection
measurement.
\end{abstract}


\maketitle

\section{Introduction}

Photonic de Broglie wavelength of a multi-photon state is the
equivalent wavelength of the whole system when all the photons in
the system act as one entity. Early work by Jacobson {\it et al}
\cite{Jac} utilized a special beam splitter that sends a whole
incident coherent state to either one of the outputs thus creating
a Sch\"odinger-cat like state. The equivalent de Broglie
wavelength in this case was shown to be $\lambda/\langle n\rangle$
with $\langle n\rangle$ as the average photon number of the
coherent state. Such a scheme can be used in precision phase
measurement to achieve the so called Heisenberg limit
\cite{hei,hol,bol,ou} of $1/\langle n\rangle$ in phase
uncertainty.

Perhaps, the easiest way to demonstrate the de Broglie wavelength
is to use the maximally entangled photon number state or the so
called NOON state of the form \cite{bol,ou,kok}
\begin{eqnarray}
|NOON\rangle = {1\over \sqrt{2}} \Big(|N\rangle_1|0\rangle_2 +
|0\rangle_1|N\rangle_2\Big), \label{1}
\end{eqnarray}
where 1,2 denote two different modes of an optical field. The $N$
photons in this state stick together either all in mode 1 or in
mode 2. Indeed, if we recombine modes 1 and 2 and make an N-photon
coincidence measurement, the coincidence rate is proportional to
\begin{eqnarray}
R_N \propto 1+\cos (2\pi N\Delta/\lambda),\label{2}
\end{eqnarray}
where $\Delta$ is the path difference between the two modes and
$\lambda$ is the single-photon wavelength. Eq.(\ref{2}) shows an
equivalent de Broglie wavelength of $\lambda/N$ for $N$ photons.

NOON state of the form in Eq.(\ref{1}) for $N=2$ case was realized
with two photons from parametric down-conversion, which led to the
demonstrations of two-photon de Broglie wavelength \cite{ou2,rar}.
For $N>2$, however, it is not easy to generate the NOON state. The
difficulty lies in the cancellation of all the unwanted terms of
$|k, N-k\rangle$ with $k\ne 0, N$ in an arbitrary N-photon
entangled state of
\begin{eqnarray}
|\Phi_N\rangle = \sum_{k=0}^N c_k|k, N-k\rangle. \label{3}
\end{eqnarray}
A number of schemes have been proposed \cite{hof,sha,liub} and
demonstrated \cite{wal,mit} which were based on some sort of
multi-photon interference scheme for the cancellation.

Without exceptions, the proposed and demonstrated schemes
\cite{hof,sha,liub,wal,mit} for the NOON state generation rely on
multi-photon coincidence measurement for revealing the phase
dependent relation in Eq.(\ref{2}). Since coincidence measurement
is a projective measurement, it may not respond to all the terms
in Eq.(\ref{3}). Indeed, Wang and Kobayashi \cite{wan} applied
this idea to a three-photon state and found that only the NOON
state part of Eq.(\ref{3}) contribute to a specially designed
coincidence measurement with asymmetric beam splitters. The
coincidence rate shows the signature dependence in the form of
Eq.(\ref{2}) on the path difference for the three-photon de
Broglie wavelength. Another projective scheme was recently
proposed by Sun {\it et al} \cite{sun1} and realized
experimentally by Resch {\it et al} \cite{res} for six photons and
by Sun {\it et al} \cite{sun2} for four photons. This scheme
directly projects an arbitrary N-photon state of the form in
Eq.(\ref{3}) onto an N-photon NOON state and thus can be scaled up
to an arbitrary N-photon case.

In this paper, we will apply the two projective schemes to the
three-photon case. The three-photon state is produced from two
pairs of photons in parametric down-conversion by gating on the
detection of one photon among them \cite{san}. We find that
because of the asymmetric beam splitters, the scheme by Wang and
Kabayashi \cite{wan} has some residual single-photon effect under
less perfect situation while the NOON state projection scheme
cancels all lower order effects regardless the situation.

The paper is organized as follows: in Sect.II, we will discuss the
scheme by Wang and Kobayashi \cite{wan} and its experimental
realization. In Sect.III, we will investigate the NOON state
projection scheme for three-photon case and implement it
experimentally. In both sections, we will deal with a more
realistic multi-mode model to cover the imperfect situations. We
conclude with a discussion.

\section{Projection by asymmetric beam splitters}

This scheme for three-photon case was first proposed by Wang and
Kobayashi \cite{wan} to use asymmetric beam splitter to cancel the
unwanted $|2,1\rangle$ or  $|1,2\rangle$ term and is a
generalization of the Hong-Ou-Mandel interferometer
\cite{ou2,rar,hon} for two-photon case. But different from the
two-photon case, the state for phase sensing is not a three-photon
NOON state since only one unwanted term can be cancelled and there
is still another one left there. So a special arrangement has to
be made in the second beam splitter to cancel the contribution
from the other term. The following is the detail of the scheme.

\subsection{Principle of Experiment}

We first start with a single mode argument by Wang and Kobayashi.
The input state is a three-photon state of
$|2\rangle_a|1\rangle_b$. The three photons are incident on an
asymmetric beam splitter (BS1) with $T\ne R$ from two sides as
shown in Fig.1. The output state can be easily found from the
quantum theory of a beam splitter as \cite{ou3,cam}
\begin{widetext}
\begin{eqnarray}
|\mathrm{BS}1\rangle_{out} = \sqrt{3T^2R}~|3_c, 0_d\rangle +
\sqrt{3TR^2}~|0_c, 3_d\rangle + \sqrt{T}(T-2R)|2_c, 1_d\rangle +
\sqrt{R}(R-2T)|1_c, 2_d\rangle.~~~~~~ \label{4}
\end{eqnarray}
\end{widetext}
When $R=2T=2/3$, the $|1_c, 2_d\rangle$ term disappears from
Eq.(\ref{4}) due to three-photon interference and Eq.(\ref{4})
becomes
\begin{eqnarray}
|\mathrm{BS}1\rangle_{out} = {\sqrt{2}\over 3} |3_c, 0_d\rangle +
{2\over 3}|0_c, 3_d\rangle - {\sqrt{3}\over 3}|2_c, 1_d\rangle .
\label{5}
\end{eqnarray}
But unlike the two-photon case, the $|2_c, 1_d\rangle$ term is
still in Eq.(\ref{5}) so that the output state is not a NOON state
of the form in Eq.(\ref{1}).

Now we can arrange a projection measurement to take out the $|2_c,
1_d\rangle$ term in Eq.(\ref{5}). Let us combine $A$ and $B$ with
another beam splitter (BS2 in Fig.1) that has same transmissivity
and reflectivity ($R=2T=2/3$) as the first BS (BS1). According to
Eq.(\ref{5}), $|2_c, 1_d\rangle$ will not contribute to the
probability $P_3(1_e,2_f)$. So only $|3_c, 0_d\rangle$ and $|0_c,
3_d\rangle$ in Eq.(\ref{5}) will contribute. The projection
measurement of $P_3(1_e,2_f)$ will cancel the unwanted middle
terms like $|2_c,1_d\rangle$ from Eq.(\ref{5}). Although the
coefficients of $|3_c, 0_d\rangle$ and $|0_c, 3_d\rangle$ in
Eq.(\ref{5}) are not equal, their contributions to $P_3(1_e,2_f)$
are the same after considering the unequal $T$ and $R$ in BS2. So
the projection measurement of $P_3(1_e,2_f)$ is responsive only to
the three-photon NOON state. Use of an asymmetric beam splitter
for the cancellation of $|2_c, 1_d\rangle$ was discussed by Sanaka
{\it et al} in Fock state filtering \cite{san}.

The above argument can be confirmed by calculating the
three-photon coincidence rate $P_3(1_e,2_f)$ directly for the
scheme in Fig.1, which is proportional to \cite{m-w8}:
\begin{eqnarray}
P_3(1_e,2_f) = \langle 2_a,1_b|\hat e^{\dag}\hat f^{\dag 2} \hat
f^2 \hat e|2_a,1_b\rangle, \label{6}
\end{eqnarray}
with
\begin{equation}
\begin{cases}
\hat e  = (\hat c + e^{i\varphi}\sqrt{2}\hat d)/\sqrt{3},\cr \hat
f = (e^{i\varphi} \hat d - \sqrt{2}\hat c)/\sqrt{3},
\end{cases}
\label{7}
\end{equation}
where we introduce a phase $\varphi$ between $A$ and $B$. But for
the first BS, we have
\begin{eqnarray}
\begin{cases}\hat c  = (\hat a + \sqrt{2}\hat b)/\sqrt{3},\cr \hat d  =
(\hat b - \sqrt{2}\hat a)/\sqrt{3}.
\end{cases}
\label{8}
\end{eqnarray}
Substituting Eq.(\ref{7}) into Eq.(\ref{6}) with Eq.(\ref{8}), we
obtain
\begin{eqnarray}
P_3(1_e,2_f) &=& \langle 2_a,1_b|\hat e^{\dag}\hat f^{\dag2}\hat
f^2\hat e|2_a,1_b\rangle \cr &=& {32\over 81} (1+\cos
3\varphi),\label{9}
\end{eqnarray}
which has a dependence  on the path difference $\Delta = \varphi
\lambda/2\pi$ that is same as in Eq.(\ref{2}) but with $N=3$.

\begin{figure}[tbp]
\includegraphics[width=8cm]{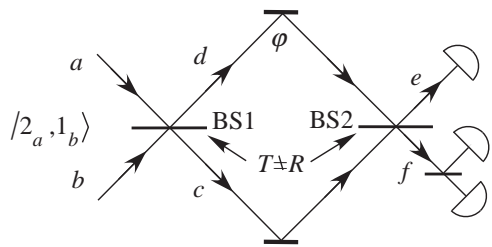}
\caption{Arrangement of asymmetric beam splitters of a
three-photon interferometer for the demonstration of the
three-photon de Broglie wavelength.}
\end{figure}

\subsection{Experiment}

Experimentally, asymmetric beam splitters are realized via
polarization projections as shown in Fig.2 and its inset (a),
where a three-photon state of $|2_H,1_V\rangle$ is incident on a
combination of two half wave plates (HWP1, HWP2) and a phase
retarder (PS). The first half wave plate (HWP1) rotates the state
$|2_H,1_V\rangle$ by an angle $\alpha $ to $|2_a,1_b\rangle$ with
\begin{eqnarray}
\begin{cases}
\hat a_H = \hat a \cos \alpha  + \hat b \sin\alpha ,\cr \hat a_V =
\hat b \cos \alpha  - \hat a \sin\alpha,
\end{cases}\label{10}
\end{eqnarray}
where $\cos \alpha = \sqrt{T} = 1/\sqrt{3}$ is the amplitude
transmissivity of the asymmetric beam splitter. Eq.(\ref{10}) is
equivalent to Eq.(\ref{7}). The phase retarder (PS) introduces the
phase shift $\varphi$ between the H and V polarization. The second
half wave plate (HWP2) makes another rotation of the same angle
$\alpha$ for the two phase-shifted polarizations and the
polarization beam splitter finishes the projection required by
Eq.(\ref{8}).

\begin{figure}
\includegraphics[width=3in]{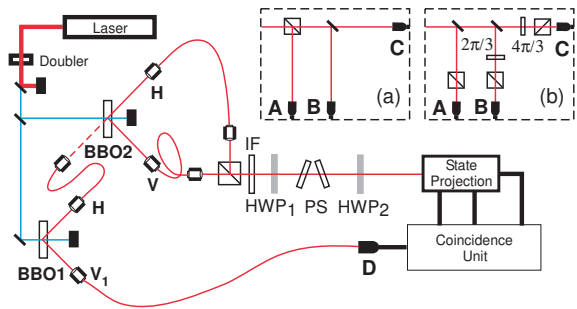}
\caption{Experimental setup. HWP1 and HWP2 is set for different
measurement, PS is the phase shifter between H-photon and
V-photon. Insets: (a) arrangement with asymmetric beam splitters
by polarization beam splitter (PBS) and (b) the three-photon NOON
state projection.}
\end{figure}

The three-photon polarization state of $|2_H,1_V\rangle$ is
prepared by using two type-II parametric down conversion processes
shown in Fig.2. This scheme was first constructed by Liu {\it et
al} \cite{liu} to demonstrate controllable temporal
distinguishability of three photons. When the delay between the
two H-photons is zero, we have the state of $|2_H,1_V\rangle$. In
this scheme, two $\beta$-Barium Borate (BBO) crystals are pumped
by two UV pulses from a common source of frequency doubled
Ti:sapphire laser operating at 780 nm. The H-photon from BBO1 is
coupled to the H-polarization mode of BBO2 while the other
V-photon is detected by detector D and serves as a trigger. The H-
and V-photons from BBO2 are coupled to single-mode fibers and then
are combined by a polarization beam splitter (PBS). The combined
fields pass through an interference filter with 3 nm bandwidth and
then go to the assembly of HWP1, PS, and HWP2 to form a
three-photon polarization interferometer. There are two schemes of
projection measurement. In this section, we deal with the first
scheme in inset (a) of Fig.2, which consists of a PBS for
projection and three detectors (A, B, C) for measuring the
quantity $P_3(1_e, 2_f)$ in Eq.(\ref{6}) by three-photon
coincidence. To realize the transformation in Eqs.(\ref{7},
\ref{8}), HWP1 and HWP2 are set to rotate the polarization by
$\alpha = \cos^{-1}(1/\sqrt{3}) = 54.7^{\circ}$. In order to
obtain an input state of $|2_H,1_V\rangle$ to the interferometer,
we need to gate the three-photon coincidence measurement on the
detection at detector D. In this way, we ensure that the two
H-photons come from two crystals separately. Otherwise, we will
have an input state of $|2_H,2_V\rangle$. The delay ($\Delta T_H$)
between the two H-photons from BBO1 and BBO2 as well as the delay
($\Delta T_V$) between the H- and V-photons are adjusted to insure
the three photons are indistinguishable in time. This is confirmed
by the photon bunching effect of the two H-photons \cite{liu} and
a generalized Hong-Ou-Mandel effect for three photons \cite{san}.

\begin{figure}[tbp]
\includegraphics[width=8cm]{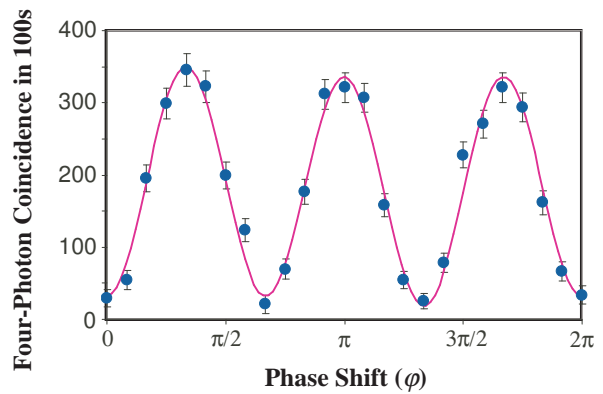}
\caption{Experimental result for projection measurement with
asymmetric beam splitters. The data is least-square-fitted to
$P_{40}(1+{\cal V}_{3}\cos 3\protect\varphi+{\cal V}_{1}\cos
\protect\varphi)$ with ${\cal V}_{3}=85\%$ and ${\cal V}_{1}=5\%$
after background subtraction. }
\end{figure}

Four-photon coincidence count among ABCD detectors is measured as
a function of the phase shift $\varphi$. The experimental result
after background subtraction is shown in Fig.3. The data is
gathered in 100 sec for each point and the error bars are one
standard deviation. Backgrounds due to three and more pairs of
photons are estimated from single and two-photon rates to
contribute 1.2/sec on average to the raw data and are subtracted.
The data clearly show a $\cos 3\varphi$ dependence except the
unbalanced minima and maxima, which indicates an extra
$\cos\varphi$ dependence. Indeed, the data fit well to the
function
\begin{eqnarray}
P_4 = P_{40}( 1+{\cal V}_3\cos 3\varphi +{\cal V}_1 \cos
\varphi)\label{11}
\end{eqnarray}
with $P_{40}=184$, ${\cal V}_3 = 85\%$ and ${\cal V}_1=5\%$. The
$\chi^2$ of the fit is 30 and is comparable to the number of data
of 25, indicating a mostly statistical cause for the error.

The appearance of the $\cos\varphi$ term in Eq.(\ref{11}) is an
indication that the cancellation of the $|2_c,1_d\rangle$ and
$|1_c,2_d\rangle$ is not complete in Eqs.(\ref{4}, \ref{9}) and
the residuals mix with the $|3_c,0_d\rangle$ and $|0_c,3_d\rangle$
terms to produce the $\cos\varphi$ term. This imperfect
cancellation is not a result of the wrong $T, R$ values but is due
to temporal mode mismatch among the three photons in the input
state of $|2_H,1_V\rangle$. To account for this mode mismatch, we
resort to a multi-mode model of the parametric down-conversion
process.

\subsection{Multi-mode analysis of three-photon interferometer with
asymmetric beam splitters}

We start by finding the multi-mode description of the quantum
state from two parametric down-conversion processes. Since the
first one serves as the input to the second one, we need the
evolution operator for the process, which was first dealt with by
Ghosh {\it et al} \cite{gho} and later by Ou \cite{ou4} and by
Grice and Walmsley \cite{wam}. In general, the unitary evolution
operator for weakly pumped type-II process is given by
\begin{eqnarray}
\hat U = 1 + \eta \int d\omega_1d\omega_2 \Phi(\omega_1,\omega_2)
\hat a_H^{\dag}(\omega_1)\hat a_V^{\dag}(\omega_2), \label{12}
\end{eqnarray}
where $\eta$ is some parameter that is proportional to the pump
strength and nonlinear coupling. For simplicity without losing
generality, we assume the two processes are identical and are
governed by the evolution operator in Eq.(\ref{12}). Furthermore,
we assume the symmetry relation $\Phi(\omega_1,\omega_2) =
\Phi(\omega_2,\omega_1)$ which is in general not satisfied but can
be achieved with some symmetrizing tricks \cite{wam2,wong}. So for
the first process, because the input is vacuum, we obtain the
output state as
\begin{widetext}
\begin{eqnarray}
|\Psi_1\rangle = \hat U |vac\rangle= |vac\rangle + \eta \int
d\omega_1d\omega_2 \Phi(\omega_1,\omega_2) \hat
a_H^{\dag}(\omega_1)\hat a_V^{\dag}(\omega_2)|vac\rangle.
\label{13}
\end{eqnarray}
The second crystal has the state of $|\Psi_1\rangle$ as its input.
So after the second crystal, the output state becomes
\begin{eqnarray}
|\Psi_2\rangle = \hat U |\Psi_1\rangle =... + \eta^2 \int
d\omega_1d\omega_2d\omega_1^{\prime}d\omega_2^{\prime}
\Phi(\omega_1,\omega_2) \Phi(\omega_1^{\prime},\omega_2^{\prime})
\hat a_H^{\dag}(\omega_1^{\prime})\hat
a_{V}^{\dag}(\omega_2^{\prime})\hat a_H^{\dag}(\omega_1)\hat
a_{V_1}^{\dag}(\omega_2)|vac\rangle, \label{14}
\end{eqnarray}
\end{widetext}
where $V_1$ and $V$ denote the two non-overlapping vertical
polarization mode from the first and second crystals,
respectively. Here we only keep the four-photon term. Although
there are other four-photon terms in the $|\Psi_2\rangle$ state
corresponding to two-pair generation from one crystal alone, they
won't contribute to what we are going to calculate. So we omit
them in Eq.(\ref{14}).

The field operators at the four detectors are given by
\begin{eqnarray}
\begin{cases}
\hat E_A(t) =\tau_1 \hat E_H(t) + \rho_1 \hat E_V(t),\cr \hat
E_B(t) = \big[\tau_2 \hat E_V(t) + \rho_2\hat
E_H(t)\big]/\sqrt{2}+... = \hat E_C(t) ,
\end{cases}
\label{15}
\end{eqnarray}
with $\tau_1=(1-2e^{i\varphi})/3, \tau_2 = (e^{i\varphi}-2)/ 3,
\rho_1= - \rho_2 =  \sqrt{2} (1+e^{i\varphi})/3$. Here we used the
equivalent relations in Eqs.(\ref{7}, \ref{8}) to establish the
connection between the field operators $\hat E_A,\hat E_B,\hat
E_C$ and $\hat E_H,\hat E_V$ and for detector D, we have
\begin{eqnarray}
\hat E_D(t) = \hat E_{V_1}(t),\label{16}
\end{eqnarray}
with
\begin{eqnarray}
\hat E_k(t) = {1\over \sqrt{2\pi}}\int d\omega \hat a_k(\omega)
e^{-i\omega t}. ~~~~~(k=H, V, V_1) \label{17}
\end{eqnarray}

The four-photon coincidence rate of ABCD is proportional to a time
integral of the correlation function
\begin{widetext}
\begin{eqnarray}
\Gamma^{(4)}(t_1,t_2,t_3,t_4) = \langle \Psi_2|\hat
E_D^{\dag}(t_4)\hat E_C^{\dag}(t_3)\hat E_B^{\dag}(t_2)\hat
E_A^{\dag}(t_1)\hat E_A(t_1)\hat E_B(t_2)\hat E_C(t_3)\hat
E_D(t_4)|\Psi_2\rangle.\label{18}
\end{eqnarray}
It is easy to first evaluate $\hat E_A(t_1)\hat E_B(t_2)\hat
E_C(t_3)\hat E_D(t_4)$:
\begin{eqnarray}
\hat E_A(t_1)\hat E_B(t_2)\hat E_C(t_3)\hat E_D(t_4) =
\big[(HHV+HVH)D\tau_1\tau_2\rho_2 + VHHD\rho_1\rho_2^2\big]/2
+...,\label{19}
\end{eqnarray}
where $H=\hat E_H$, $V= \hat E_V$, $D= \hat E_D$ for short and we
keep the time ordering of $t_1t_2t_3$. We also drop five terms
that give zero result when they operate on $|\Psi_2\rangle$. It is
now straightforward to calculate the quantity $\hat E_A(t_1)\hat
E_B(t_2)\hat E_C(t_3)$ $\hat E_D(t_4)|\Psi_2\rangle$, which has
the form of
\begin{eqnarray}
&&\hat E_A(t_1)\hat E_B(t_2)\hat E_C(t_3)\hat
E_D(t_4)|\Psi_2\rangle = {\eta^2\over 2}
\Big\{\big[G(t_1,t_2,t_3,t_4)+G(t_2,t_1,t_3,t_4)+
G(t_1,t_3,t_2,t_4)+ G(t_3,t_1,t_2,t_4)\big]\tau_1\tau_2\rho_2
\cr&&\hskip 3in
 + \big[G(t_2,t_3,t_1,t_4)+G(t_3,t_2,t_1,t_4)\big]
\rho_1\rho_2^2\Big\}|vac\rangle,\label{20}
\end{eqnarray}
where
\begin{eqnarray}
G(t_1,t_2,t_3,t_4)= g(t_1,t_3)g(t_2,t_4)~~~~{\rm
with}~~~g(t,t^{\prime})\equiv {1\over 2\pi}\int d\omega_1d\omega_2
\Phi(\omega_1,\omega_2) e^{-i(\omega_1t +\omega_2t^{\prime})}.
\label{21}
\end{eqnarray}
Substituting Eq.(\ref{20}) into Eq.(\ref{18}) and carrying out the
time integral, we obtain
\begin{eqnarray}
P_4 &\propto& \int dt_1dt_2dt_3dt_4
\Gamma^{(4)}(t_1,t_2,t_3,t_4)\cr &=&{|\eta|^4\over 4}\int
d\omega_1d\omega_2d\omega_1^{\prime}d\omega_2^{\prime}
\Big|\big[\Phi(\omega_1,\omega_2)
\Phi(\omega_1^{\prime},\omega_2^{\prime})+\Phi(\omega_1,\omega_1^{\prime})
\Phi(\omega_2,\omega_2^{\prime})\big](\tau_1\tau_2\rho_2+\rho_1\rho_2^2)\cr
&&\hskip 3in+2\Phi(\omega_1^{\prime},\omega_2)
\Phi(\omega_1,\omega_2^{\prime})\tau_1\tau_2\rho_2\Big|^2.
\label{22}
\end{eqnarray}
\end{widetext}
With $\tau_1,\rho_1,\tau_2,\rho_2$, we can further reduce
Eq.(\ref{22}) to
\begin{eqnarray}
P_4\propto {2|\eta|^4(17{\cal A}+7{\cal E})\over 243}\Big[1+{\cal
V}_3 \cos 3\varphi +{\cal V}_1\cos \varphi\Big] \label{23}
\end{eqnarray}
where
\begin{eqnarray}
{\cal V}_3 = {8({\cal A}+2{\cal E})\over 17{\cal A} + 7{\cal E}},
~~~ {\cal V}_1 = {9({\cal A}-{\cal E})\over 17{\cal A} + 7{\cal
E}}.\label{24}
\end{eqnarray}
and
\begin{eqnarray}
&&{\cal A} =\int d\omega_1d\omega_2d\omega_1^{\prime}
d\omega_2^{\prime} |\Phi(\omega_1,\omega_2)
\Phi(\omega_1^{\prime},\omega_2^{\prime})|^2 \label{25}\\ &&{\cal
E} =\int d\omega_1d\omega_2d\omega_1^{\prime} d\omega_2^{\prime}
\Phi^*(\omega_1,\omega_2)
\Phi^*(\omega_1^{\prime},\omega_2^{\prime})\cr&&\hskip 1.5in
\times \Phi(\omega_1^{\prime},\omega_2)
\Phi(\omega_1,\omega_2^{\prime}). ~~~~~\label{26}
\end{eqnarray}
In deducing Eqs.(\ref{22}--\ref{26}), we used the symmetry
relation $\Phi(\omega_1,\omega_2)=\Phi(\omega_2,\omega_1)$.

Obviously, when ${\cal A}={\cal E}$, Eq.(\ref{23}) completely
recovers to Eq.(\ref{9}). In practice, we always have ${\cal
A}\ge{\cal E}$ by Schwartz inequality. When ${\cal E} < {\cal A}$,
Eq.(\ref{23}) has the same form as Eq.(\ref{11}), indicating that
the multi-mode analysis indeed correctly predicts the imperfect
cancellation of the $|2_H,1_V\rangle$ and $|2_V,1_H\rangle$ terms
in Eqs.(\ref{4}, \ref{9}). If we use the experimentally measured
${\cal V}_3$ and ${\cal V}_1$ in Eq.(\ref{24}), we will obtain two
inconsistent values of ${\cal E}/{\cal A}$: $({\cal E}/{\cal
A})_3=0.65$ and $({\cal E}/{\cal A})_1=0.87$. The discrepancy is
the result of the break up of the symmetry relation of
$\Phi(\omega_1,\omega_2) = \Phi(\omega_2,\omega_1)$ for type-II
parametric down-conversion, which is reflected in the
less-than-unit visibility of the two-photon interference. This
imperfection can be modelled as spatial mode mismatch and
approximately modifies Eq.(\ref{24}) as
\begin{eqnarray}
{\cal V}_3 = v_1^3{8({\cal A}+2{\cal E})\over 17{\cal A} + 7{\cal
E}}, ~~~ {\cal V}_1 = v_1 {9({\cal A}-{\cal E})\over 17{\cal A} +
7{\cal E}},\label{27}
\end{eqnarray}
where $v_1$ is the equivalent reduced visibility in single-photon
interference due to spatial mode mismatch. With the extra
parameter $v_1$ in Eq.(\ref{27}), we obtain a consistent $({\cal
E}/{\cal A})=0.86$ with $v_1 = 0.96$.

\section{NOON state projection measurement}

The projection measurement discussed in the previous section
relies on the cancellation of some specific terms and therefore
cannot be applied to an arbitrary photon number. In the following,
we will discuss another projection scheme that can cancel all the
unwanted terms at once and thus can be scaled up.

\subsection{Principle of experiment}

The NOON-state projection measurement scheme was first proposed by
Sun {\it et al} \cite{sun1} and realized by Resch {\it et al}
\cite{res} for six-photon case and by Sun {\it et al} \cite{sun2}
for the four-photon case. Since it is based on a multi-photon
interference effect, it was recently used to demonstrate the
temporal distinguishability of an N-photon state \cite{xia,ou5}.
Here we will apply it to a three-photon superposition state for
the demonstration of three-photon de Broglie wave length without
the NOON state.

The NOON-state projection measurement scheme for three-photon case
is sketched in inset (b) of Fig.1. In this scheme, the input field
is first divided into three equal parts. Then each part passes
through a phase retarder that introduces a relative phase
difference of $0, 2\pi/3,4\pi/3$ respectively between the H- and
V-polarization. The phase shifted fields are then projected to
$135^{\circ}$ direction by polarizers before being detected by A,
B, C detectors, respectively. It was shown that the three-photon
coincidence rate is proportional to
\begin{equation}
P_3 \propto {1\over 18}\Big|\langle
NOON_3|\Phi_3\rangle\Big|^2,\label{28}
\end{equation}
where  $|NOON_3\rangle = (|3_H,0_V\rangle
-|0_H,3_V\rangle)/\sqrt{2}$ and $|\Phi_3\rangle =
c_0|3_H,0_V\rangle + c_1|2_H,1_V\rangle + c_2|1_H,2_V\rangle +
c_3|0_H,3_V\rangle$. Note that since $|2,1\rangle ,|1,2\rangle$
are orthogonal to the NOON-state, their contributions to $P_3$ are
zero. Assuming that $|c_0|=|c_3|=c$ and there is a relative phase
of $\varphi$ between H and V so that $c_0=c, c_3=ce^{i3\varphi}$,
we obtain from Eq.(\ref{28})
\begin{equation}
P_3 \propto {|c|^2\over 18}(1-\cos 3\varphi),\label{29}
\end{equation}
which is exactly in the form of Eq.(\ref{2}) with $N=3$, showing
the three-photon de Broglie wave length.

\subsection{Experiment}

From Sect.IIB, we learned that a state of $|2_H,1_V\rangle$ can be
produced with two parametric down-conversion processes. This state
will of course give no contribution to the NOON state projection
since it is orthogonal to the NOON state. On the other hand, we
can rotate the state by 45$^{\circ}$. Then the state becomes
\cite{ou3,cam}
\begin{eqnarray}
&&|\Phi_3\rangle = \sqrt{3\over 8}\Big(|3_H,0_V\rangle
-|0_H,3_V\rangle\Big) \cr &&\hskip 1in +
{1\over\sqrt{8}}\Big(|2_H,1_V\rangle -
|1_H,2_V\rangle\Big),~~~~~~\label{30}
\end{eqnarray}
which has the NOON state component with $c=\sqrt{3/8}$.

Experimentally, the three-photon state of $|2_H,1_V\rangle$ is
prepared in the same way described in Sect.IIB and shown in Fig.2.
Different from Sect.IIB, the polarizations of the prepared state
are rotated 45$^{\circ}$ by HWP1 to achieve the state in
Eq.(\ref{30}). The phase shifter (PS) then introduces a relative
phase difference $\varphi$ between the H and V polarizations and
HWP2 is set at zero before the NOON state projection measurement
is performed [Inset (b) of Fig.2]. As before, a four-photon
coincidence measurement among ABCD detectors is equivalent to a
three-photon coincidence measurement by ABC detectors gated on the
detection at D, which is required for the production of
$|2_H,1_V\rangle$.

\begin{figure}[tbp]
\includegraphics[width=8cm]{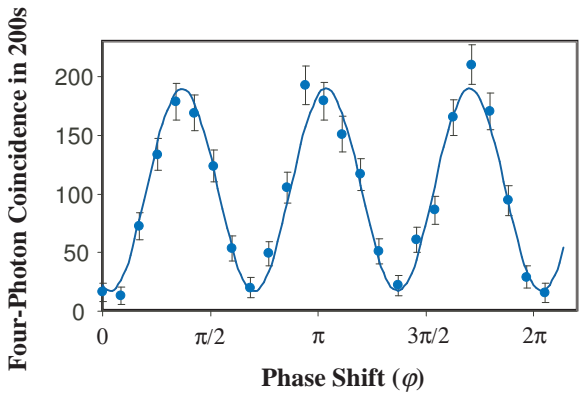}
\caption{Experimental result for the NOON state projection
measurement. The data is least-square-fitted to $P_{40}(1+{\cal
V}_{3}\cos 3\protect\varphi$ with ${\cal V}_{3}=84\%$ after
background subtraction.}
\end{figure}

Four-photon coincidence count among ABCD detectors is registered
in 200 sec as a function of the phase $\varphi$ (PS). The data
after subtraction of background contributions is plotted in Fig.4.
It clearly shows a sinusoidal dependence on $\varphi$ with a
period of $2\pi/3$. The solid curve is a chi-square fit to the
function of $P_4=P_{40}[1+{\cal V}_3\cos 3(\varphi+\varphi_0)]$
with $P_{40}=103/200$sec and ${\cal V}_3=0.84$. The $\chi^2$ of
the fit is 24.3, which is comparable to the number of data of 25
indicating a good statistical fit.

The less-than-unit visibility is a result of temporal
distinguishability among the three photons produced from two
crystals. It can only be accounted for with a multi-mode model of
the state given in Sect.IIC. Let's now apply it to the current
scheme.

\subsection{Multi-mode analysis}

The input state is same as Eq.(\ref{14}). But the field operators
are changed to
\begin{eqnarray}
\begin{cases}
\hat E_A(t) = \big[\hat E_+(t) - e^{i\varphi}\hat
E_-(t)\big]/\sqrt{6} + ...,\cr \hat E_B(t) = \big[\hat E_+(t)
-e^{i\varphi} \hat E_-(t)e^{i2\pi/3}\big]/\sqrt{6} + ...,\cr \hat
E_C(t) = \big[\hat E_+(t) - e^{i\varphi}\hat
E_-(t)e^{i4\pi/3}\big]/\sqrt{6} + ...,
\end{cases}
\label{31}
\end{eqnarray}
with
\begin{eqnarray}
\begin{cases}
\hat E_+(t) = \big[\hat E_H(t) + \hat E_V(t)\big]/\sqrt{2},\cr
\hat E_-(t) = \big[\hat E_H(t) - \hat E_V(t)\big]/\sqrt{2},
\end{cases}
\label{32}
\end{eqnarray}
where we omit the vacuum input fields and $\varphi$ is the phase
shift introduced by PS in Fig.2. The field operator for detector D
is same as Eq.(\ref{16}).

As in Sect.IIC, the four-photon coincidence rate is related to the
correlation function in Eq.(\ref{18}) and we can first evaluate
$\hat E_A(t_1)\hat E_B(t_2)\hat E_C(t_3)\hat E_D(t_4)$. With the
field operators in Eq.(\ref{31}), we obtained
\begin{eqnarray}
&&\hat E_A(t_1)\hat E_B(t_2)\hat E_C(t_3)\hat E_D(t_4) \cr
&&\hskip 0.3in = \big(HHV a_1 + HVH a_2 +VHH
a_3\big)/12\sqrt{12},~~~~~\label{33}
\end{eqnarray}
with
\begin{eqnarray}
\begin{cases} a_1=1+ e^{i3\varphi}+ 2e^{i(2\varphi+2\pi/3)}+
2e^{i(\varphi+4\pi/3)}\cr a_2 = 1+ e^{i3\varphi}+
2e^{i(2\varphi+4\pi/3)}+ 2e^{i(\varphi+2\pi/3)}\cr a_3=1+
e^{i3\varphi}+ 2e^{i2\varphi}+ 2e^{i\varphi},\label{34}
\end{cases}
\end{eqnarray}
where the notations are same as in Eq.(\ref{19}) and we used the
identity $1+e^{i2\pi/3}+e^{i4\pi/3} = 0$. As before, we also drop
five terms that give zero result when they operate on
$|\Psi_2\rangle$. Now we can calculate the quantity $\hat
E_A(t_1)\hat E_B(t_2)\hat E_C(t_3)\hat E_D(t_4)|\Psi_2\rangle$,
which has the form of
\begin{widetext}
\begin{eqnarray}
\hat E_A(t_1)\hat E_B(t_2)\hat E_C(t_3)\hat E_D(t_4)|\Psi_2\rangle
= {\eta^2\over \sqrt{12^3}}
\Big\{\big[G(t_1,t_2,t_3,t_4)+G(t_2,t_1,t_3,t_4)\big] a_1+\big[
G(t_1,t_3,t_2,t_4)+ G(t_3,t_1,t_2,t_4)\big]a_2\nonumber
\end{eqnarray}
\begin{eqnarray}
\hskip 3in
 + \big[G(t_2,t_3,t_1,t_4)+G(t_3,t_2,t_1,t_4)\big]
a_3\Big\}|vac\rangle,\label{35}
\end{eqnarray}
\end{widetext}
where $G(t_1,t_2,t_3,t_4)$ is given in Eq.(\ref{21}). After the
time integral, we obtain
\begin{eqnarray}
P_4(NOON) \propto {|\eta|^4(2{\cal A}+{\cal E})\over
72}\big(1+{\cal V}_3 \cos 3\varphi \big) \label{36}
\end{eqnarray}
with
\begin{eqnarray}
{\cal V}_3(NOON) = {{\cal A} +2{\cal E}\over 2{\cal A}+{\cal E}},
\label{37}
\end{eqnarray}
where ${\cal A}$ and ${\cal E}$ are given in Eqs.(\ref{25},
\ref{26}). Note that the terms such as $\cos2\varphi, \cos\varphi$
are absent in Eq.(\ref{37}) even in the non-ideal case of ${\cal
E}<{\cal A}$. This is because of the symmetry among the three
detectors A, B, C involved in the three-photon NOON state
projection measurement. When spatial mode mismatch is considered,
the visibility is changed to
\begin{eqnarray}
{\cal V}_3(NOON) = v_1^3{{\cal A} +2{\cal E}\over 2{\cal A}+{\cal
E}}. \label{38}
\end{eqnarray}
With $v_1$ and the quantity ${\cal E}/{\cal A}$ obtained in
Sect.IIC, we have ${\cal V}_3(NOON) =0.85$, which is close to the
observed value of 0.84 in Sect.IIIB.

\section{Summary and Discussion}

In summary, we demonstrate a three-photon de Broglie wavelength by
using two different schemes of projection measurement without the
need for a hard-to-produce NOON state. Quantum interference is
responsible for the cancellation of the unwanted terms. The first
scheme by asymmetric beam splitters targets specific terms while
the second one by NOON state projection cancels all the unwanted
terms at once. We use a multi-mode model to describe the non-ideal
situation encountered in the experiment and find good agreements
with the experimental results.

Since the scheme by asymmetric beam splitters is only for some
specific terms, it cannot be easily scaled up to arbitrary number
of photons although the extension to the four-photon case is
available. The extension of the scheme by NOON state projection to
arbitrary number of photons is straightforward. In fact,
demonstrations with four and six photons have been done with
simpler states \cite{res,sun2}.

On the other hand, the scheme of NOON state projection need to
divide the input fields into $N$ equal parts while the scheme with
asymmetric beam splitters requires less partition. So the latter
will have higher coincidence rate than the former. In fact, Fig.3
and Fig.4 show a ratio of 4 after pump intensity correction. This
is consistent with the ratio of 4.8 from Eqs.(\ref{23}, \ref{36})
when ${\cal E= A}$. The difference may come from the different
collection geometry in the layout.

The dependence of the visibility in  Eqs.(\ref{27}, \ref{37}) on
the quantity ${\cal E/A}$ reflects the fact that the interference
effect depends on the temporal indistinguishability of the three
photons. From previous studies \cite{ou6,tsu,ou7,sun1,liu}, we
learned that the quantity ${\cal E/A}$ is a measure of
indistinguishability between two pairs of photons in parametric
down-conversion. In our generation of the $|2_H, 1_V\rangle$
state, one of the H-photon is from another pair of down-converted
photons. So to form an indistinguishable three-photon state, we
need pair indistinguishability, i.e., ${\cal E/A}\rightarrow 1$.

\begin{acknowledgments}
This work was funded by National Fundamental Research Program of
China (2001CB309300), the Innovation funds from Chinese Academy of
Sciences, and National Natural Science Foundation of China (Grant
No. 60121503 and No. 10404027)). ZYO is also supported by the US
National Science Foundation under Grant No. 0245421 and No.
0427647.
\end{acknowledgments}

\end{document}